\documentclass[a4paper]{jpconf}
\usepackage{graphicx}
\graphicspath{}
\usepackage[numbers,sort&compress]{natbib}

\begin{document}

\title{Charged excitons or trions in 2D parabolic quantum dots?}

\author{Nguyen Hong Quang$^{1,2}$ and Nguyen Que Huong$^{3}$}
\address{$^1$ Graduate University of Science and Technology, VAST, \\\hspace{0.1cm} 18 Hoang Quoc Viet, Nghia Do, Cau Giay, Hanoi, Vietnam\\[0.2cm]}
\address{$^2$ Institute of Physics, Vietnam Academy of Science and Technology (VAST), \\\hspace{0.1cm} 18 Hoang Quoc Viet, Nghia Do, Cau Giay, Hanoi, Vietnam\\[0.2cm]}
\address{$^3$ Marshall University, One John Marshall Drive, Huntington WV 25701}

\ead{nhquang@iop.vast.vn}

\begin{abstract}
So far in the literature the terms ``charged exciton" and ``trion" are often confused with each other and mostly considered as the same. In this work we show this is not the case in 2D quantum dots with a parabolic confinement. By using the unrestricted Hartree-Fock method the energy and binding energy of both charged excitons and trions in 2D parabolic quantum dots are calculated in dependence on the confinements of charge carriers in quantum dot. It is shown that the binding energies of the charged exciton and the trion behave differently in regard to the ratios of the confinements between the electron and hole. The effect of the external magnetic field on the binding energies of charged excitons has been also considered.
\\[0.5cm]
\end{abstract}
\section{Introduction}
Exciton effects have been intensively studied in  low-dimensional semiconductor systems not only theoretically but experimentally in recent years due to their critical role in electronics, optoelectric devices and quantum information technology in the future \cite{Bimberg,Wang,Peter}. 

A trion is a bound state of three charged particles, which can be either two electrons associated with a hole or two holes associated with an electron. Trions can also be bound states of an exciton with an electron or an exciton with a hole. In this case they are also known as negatively charged excitons or positively charged excitons \cite{Lampert,Kheng,Climente,Ganchev}. In fact, in trion states, like excitons and other exciton-based systems, such as biexcitons, made up of pairs of electrons - holes, the charge carriers are correlated with each other through Coulomb interaction, which has been studied a lot before, but has attracted great attention recently, especially in
atomic-size thin semiconductor layers based on transition metal dichalcogenide monolayers \cite{Mak,Ross,Glazov,Lan}.

Quite recently, new terms ``duo", ``trio", ``quatuor" for quasi-particle consisting of two, three and four charge carriers, respectively has been introduced in quantum dots by Combescot \cite{Combescot} and “quadron” for four charge carriers in 2D semiconductor quantum dots by us \cite{Quang1,Quang2,Quang3} to distinguish them from conventional excitons, charged excitons and biexcitons. 
In particular, self-organized semiconductor quantum dots are still in special interest of researchers today, since the binding energies of quasi-particles in the system are large and the effect of magnetic fields can be studied along with the symmetry of the quantum dot. \cite{Combescot,Quang1,Quang2,Quang3,Rodt,Bester,Schliwa,Tsai,Abbarchi,Zielinski,Quang}. 

In \cite{Combescot,Quang1,Quang2,Quang3} the authors have demonstrated the essential differences between ordinary biexciton and  “quatuor” \cite{Combescot} or ``quadron" \cite{Quang1,Quang2,Quang3} in quantum dots. 
The main reason for these differences is the confinement impact on the charged carriers in the quantum dot, where multi-particle pair interactions between four carriers should be considered equally. 
In our previous works \cite{Quang1,Quang2,Quang3} we have demonstrated that the confinement and the Coulomb interaction in a small InAs quantum dot lead to a bound quadron, rather a conventional biexciton. 

To develop further the investigations on this direction, in this work we consider differences between usual charged excitons and trions in self-assembled semiconductor quantum dots with 2D parabolic potential of confinement. By using the Hatree-Fock method, the energies and binding energies of the charged excitons and trions have been calculated for an entire range of confinement ratio between the electron and the hole. We will show that the binding energies of the charged exciton and the trion behave differently in regard to the ratios of the confinements between the electron and hole. The magnetic field effect on the binding energy of the charged excitons has been also presented. 

\section{The model}
We consider the system of electrons and holes in a 2D quantum dot with parabolic potential, and in a magnetic field $\vec{B} \| z$. The total Hamiltonian of the system of $N$ electrons and $M$ holes ($N=2,M=1$ or $N=1,M=2$ for charged excitons and trions)  can be written in the effective-mass approximation, as follows \cite{Quang1,Quang2,Quang3}, 
\begin{equation}
\widehat{H} = \sum_{i=1}^{N}h(\vec{r}_i) +\sum_{k=1}^{M}h'(\vec{r}_k) + \sum_{i=1; i<j}^{N}
\frac{e^2}{\epsilon r_{ij}} +\sum_{k=1; k<l}^{M}\frac{e^2}{\epsilon r_{kl}}-\sum_{i=1}^{N}\sum_{k=1}^M\frac{e^2}{\epsilon r_{ik}} \ ,
\end{equation}
 where $h(\vec{r}_i)$  and  $h'(\vec{r}_k)$ are the Hamiltonians of a single electron and a single hole, given by the formula (\ref{eq2}) and (\ref{eq3}) and the other terms describe the electron-electron, the hole-hole and the electron-hole Coulomb interactions, respectively, and  $\epsilon$ is the material's dielectric constant. 

For a single electron and a single hole in 2D parabolic quantum dots and in the presence of a magnetic field, the Hamiltonians are:
\begin{eqnarray}\label{eq2}
h(\vec{r}_i) &=& -\frac{\nabla^2_i}{2 m^*_e}
+\frac{m^*_e}{2}(\omega^2_e+\frac{1}{4}\omega^2_{ce})r^2_i
+\frac{1}{2}\omega_{ce} \hat{L}_{zi} \ ,\\[0.3cm] 
h'(\vec{r}_k) &=& -\frac{\nabla^2_k}{2 m^*_h}
+\frac{m^*_h}{2}(\omega^2_h+\frac{1}{4}\omega^2_{ch})r^2_k
+\frac{1}{2}\omega_{ch} \hat{L}_{zk} \ , \label{eq3}
\end{eqnarray}
where the effective mass of the electron (hole) are denoted by $m^*_e$ ($m^*_h$) and their confinement potentials are denoted  by  $\omega_e$ ($\omega_h$). The expressions for the cyclotron frequency for the electron (hole) are $ \omega_{ce} = eB/m^*_e$ ($ \omega_{ch} = eB/m^*_h$). Note that in (\ref{eq2}) and (\ref{eq3}) we neglect the Zeeman effect because of its smallness. The last term relates with the orbital angular momentum of the electron (hole), respectively $\hat{L}_{zi}$ ($\hat{L}_{zk}$).

In polar coordinates $\vec{r}=(r,\varphi)$ we have the single electron's eigenfunction and the eigenvalue in the states $(n,m)$, as follows:
\begin{eqnarray}\label{eq4}
    \chi^e_{n,m}(r,\varphi)&=&\frac{1}{\sqrt{2\pi}}e^{i m \varphi}\sqrt{\frac{2 n!}{(n+|m|)!}}\alpha_e(\alpha_e r)^{|m|} e^{-(\alpha_e r)^2/2} L_n^{|m|}((\alpha_e r)^2) \ , \label{eq:chie}\\[0.3cm]
E^e_{n,m}&=&\Omega_e(2n + | m| +1)+\frac{1}{2}m\omega_{c_e} \ , \label{eq5}
\end{eqnarray}
where 
$\alpha_e = \sqrt{m_e^* \Omega_e}, \Omega_e = (\omega_e^2+\frac{1}{4}\omega_{c_e}^2)^{1/2}$, and $L_n^{|m|}(r)$ is generalized Laguerre polynomial. Note that we have the analogous formulae for the  hole, in which the index $h$ for the hole replaces the index $e$ for the electron.

In \cite{Quang1,Quang2,Quang3} we have presented in details the Hartree-Fock-Roothaan formulation for the multi-particle systems, using the functions (\ref{eq4}) as  basic functions, here we only show the final expression for the total energy of the system in our study:
\begin{eqnarray}\label{eq:energy}
E&=&\frac{1}{2}\sum_{\mu,\nu}\Big\{\delta_{\mu\nu}P_{\mu\nu}^T
[\Omega_e(2n+|m|+1)+m\omega_{c_e}] +P_{\mu\nu}^\alpha F_{\nu\mu}
^\alpha + P_{\mu\nu}^\beta F_{\nu\mu}^\beta\Big\}  \nonumber\\
&+&\frac{1}{2}\sum_{\mu,\nu}\Big\{\delta_{\mu\nu}P_{\mu \nu}'^{\ T}[\Omega_h(2n+|m|+1)-m\omega_{c_h}] + P_{\mu\nu}'^{\ \alpha} F_{\nu\mu}'^{\ \alpha} + P_{\mu\nu}'^{\ \beta} F_{\nu\mu}'^{\ \beta}\Big\} , 
\end{eqnarray}
where the explicit expressions for 
$ P_{\mu \nu}^T , P_{\mu \nu}'^{\ T}$ , $P_{\mu \nu}^\alpha , P_{\mu \nu}^\beta$ , $P_{\mu \nu}'^{\ \alpha}$ , $P_{\mu \nu}'^{\ \beta}$
and 
$F_{\mu\nu}^{\alpha, \beta}, 
F_{\mu\nu}'^{\ \alpha, \beta}$ are given in \cite{Quang1}.

\section{The results and discussions}
For numerical calculations, we take the parameters for InAs/GaAs quantum dots with $m^*_e = 0.067 m_o$,\  $\omega_e = 49$\ meV, \  $m^*_h = 0.25 m_o$,\  $\omega_h = 25$\ meV,$\  \epsilon_s = 12.53$. 
The units of length is $a_B^* = \epsilon_s/m^*_e e^2 = 9.9$\ nm, and the units of energy are is $2Ry^* = m^*_e e^4/\epsilon^2_s = 11.61$\ meV \cite{Quang1,Quang2,Quang3}. 

In the study of the effect of confinement on the binding energies, for the reason of comparison we use two parameter sets, namely set 1 with 
$\omega_e = 49$\ meV, $m^*_e = 0.067 m_o$, and \  $m^*_h = 0.25 m_o$, and set 2 with $\omega_e = 49$\ meV, $m^*_e = 0.067 m_o$, and \  $m^*_h = 0.067 m_o$.

Denote $m_e = m^*_e /m_o; m_h=m^*_h/m_o$, we have for the set 1: $m_e/m_h =0.27$, and for the set 2: $m_e/m_h =1$.

\begin{figure}[h]
\begin{center}
\begin{minipage}{6cm}
\includegraphics[width=6cm]{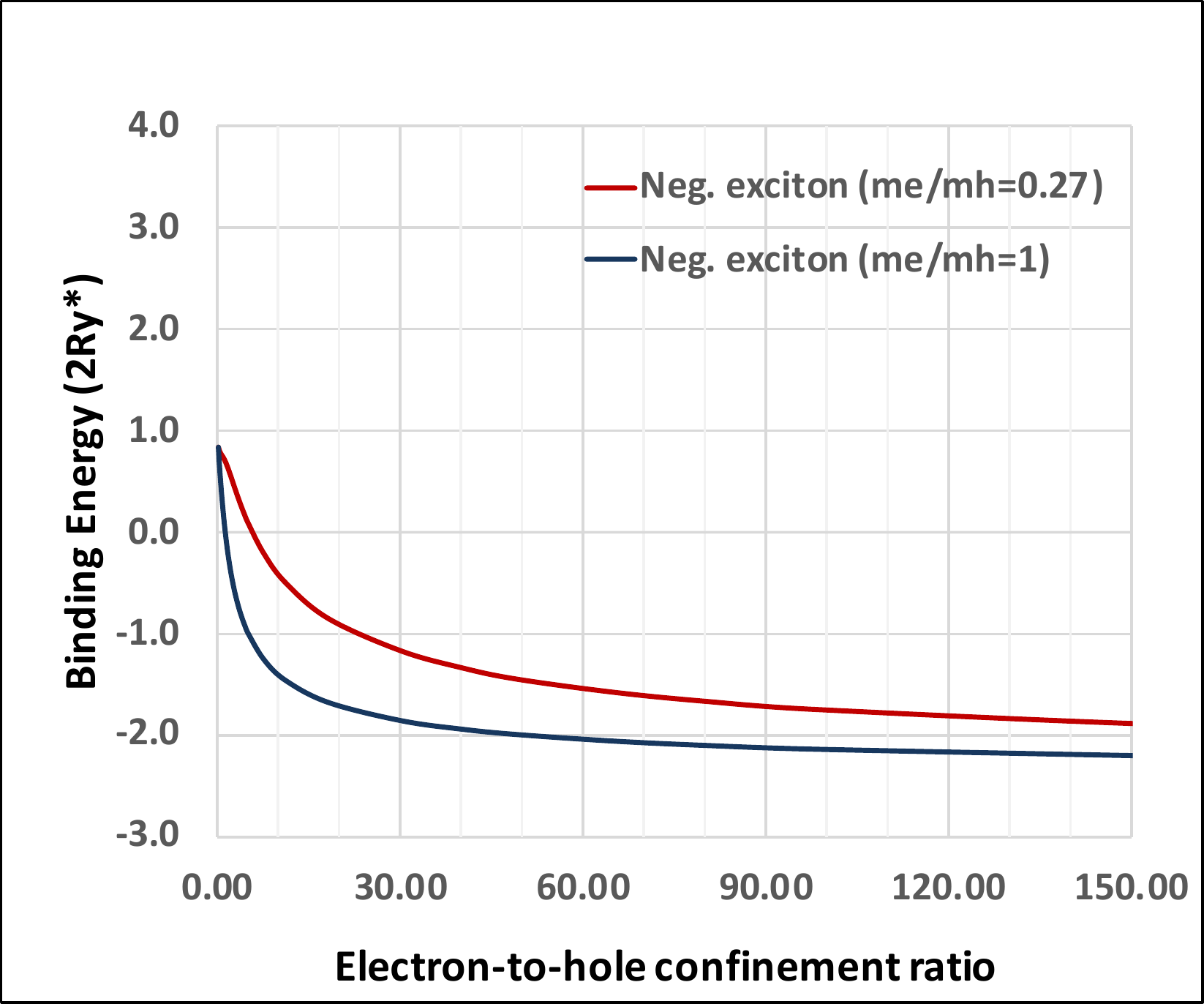}
\caption{\label{Fig1}(color online) The comparison of the negatively charged exciton binding energies with the $m_e/m_h=0.27$ and $m_e/m_h=1$  as function of electron-to-hole confinement.}
\end{minipage}\hspace{2pc}%
\begin{minipage}{6cm}
\includegraphics[width=6cm]{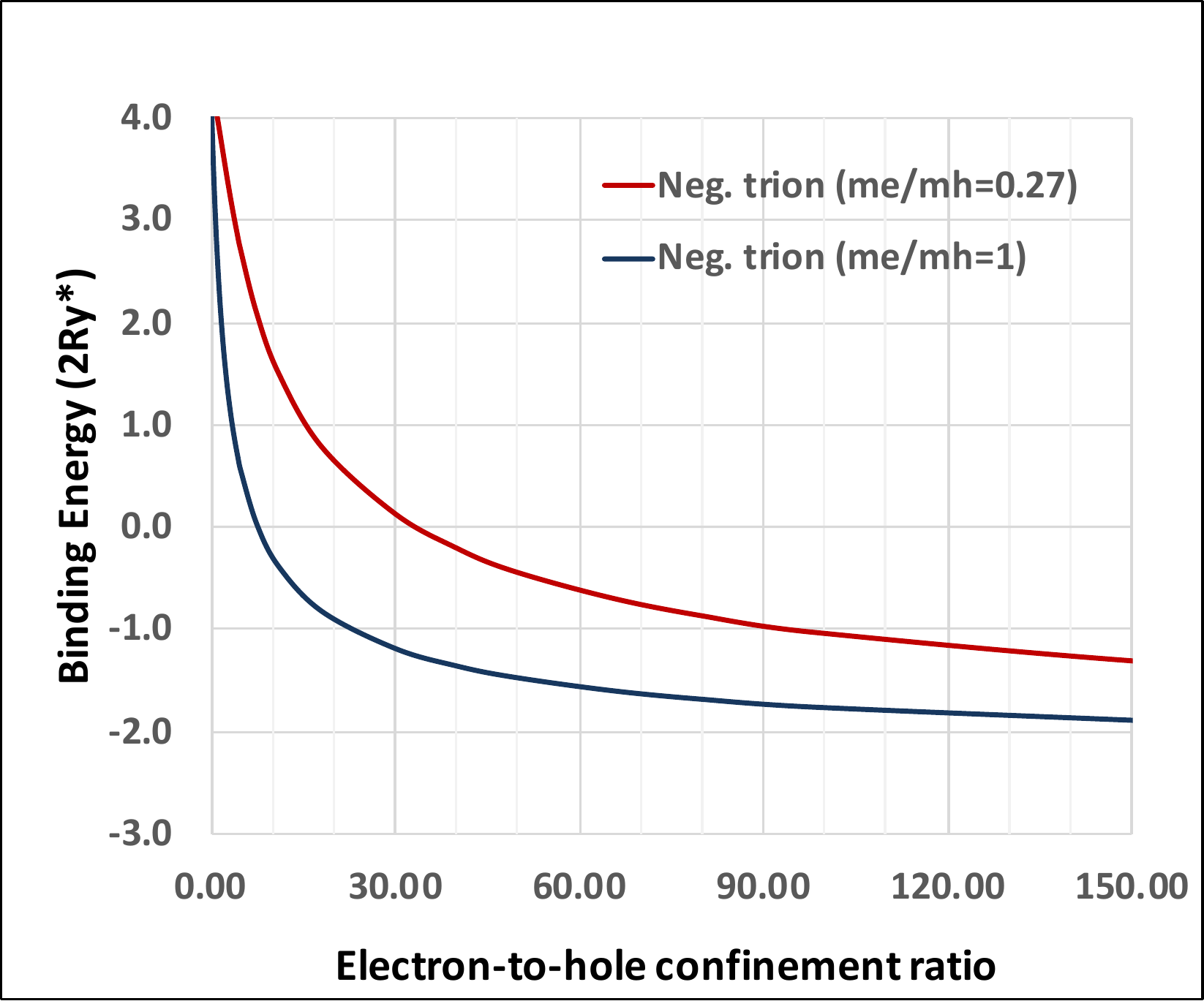}
\caption{\label{Fig2}(color online) The comparison of the negative trion binding energies with  the $m_e/m_h=0.27$ and $m_e/m_h=1$ as function of electron-to-hole confinement ratio.\\}
\end{minipage} 
\end{center}
\end{figure}
\begin{figure}[h!]
\begin{center}
\begin{minipage}{6cm}
\includegraphics[width=6cm]{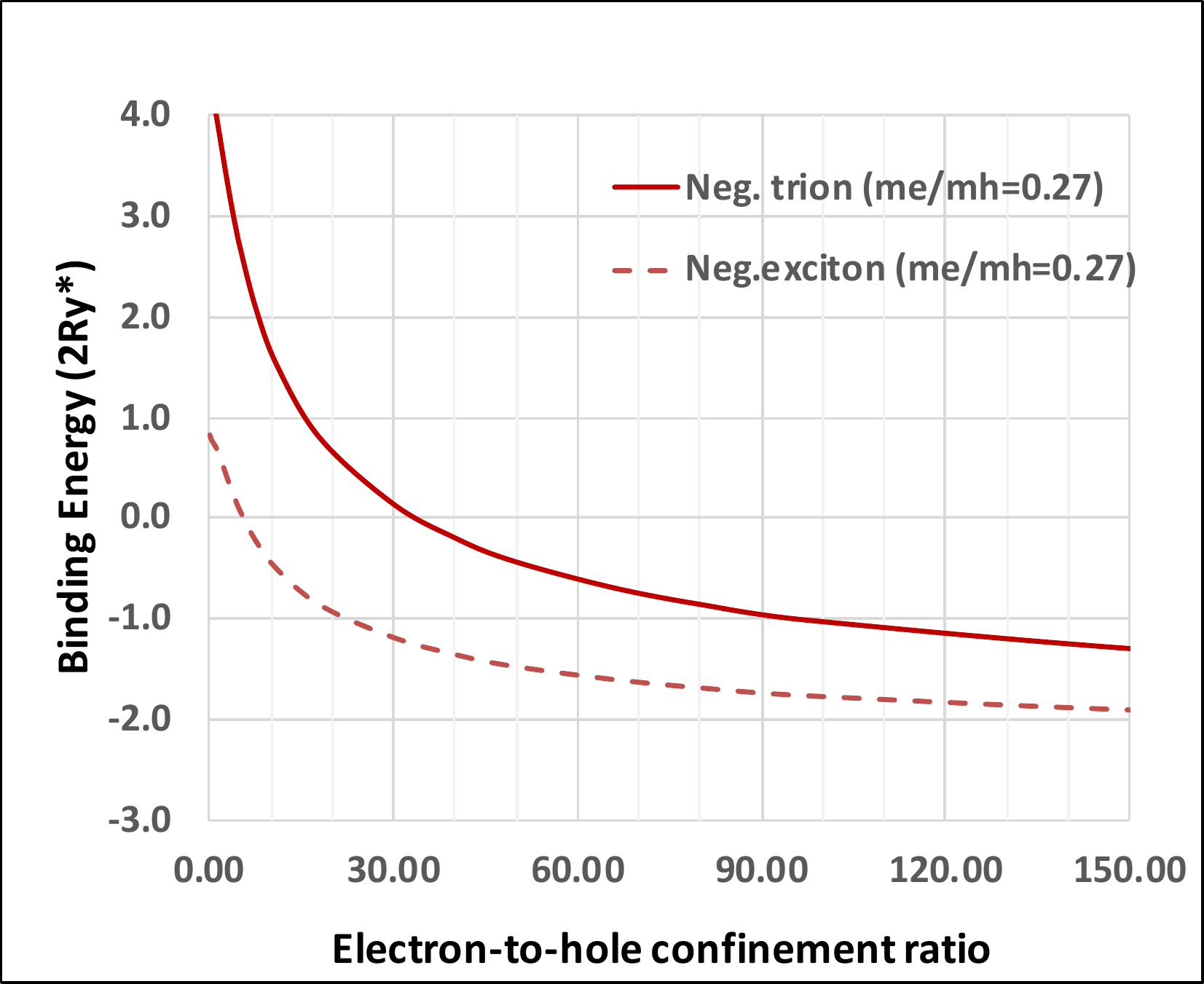}
\caption{\label{Fig3}(color online) The comparison of the negative trion and negatively charged exciton binding energies with the $m_e/m_h=0.27$ as function of electron-to-hole confinement ratio.}
\end{minipage}\hspace{2pc}%
\begin{minipage}{6cm}
\includegraphics[width=6cm]{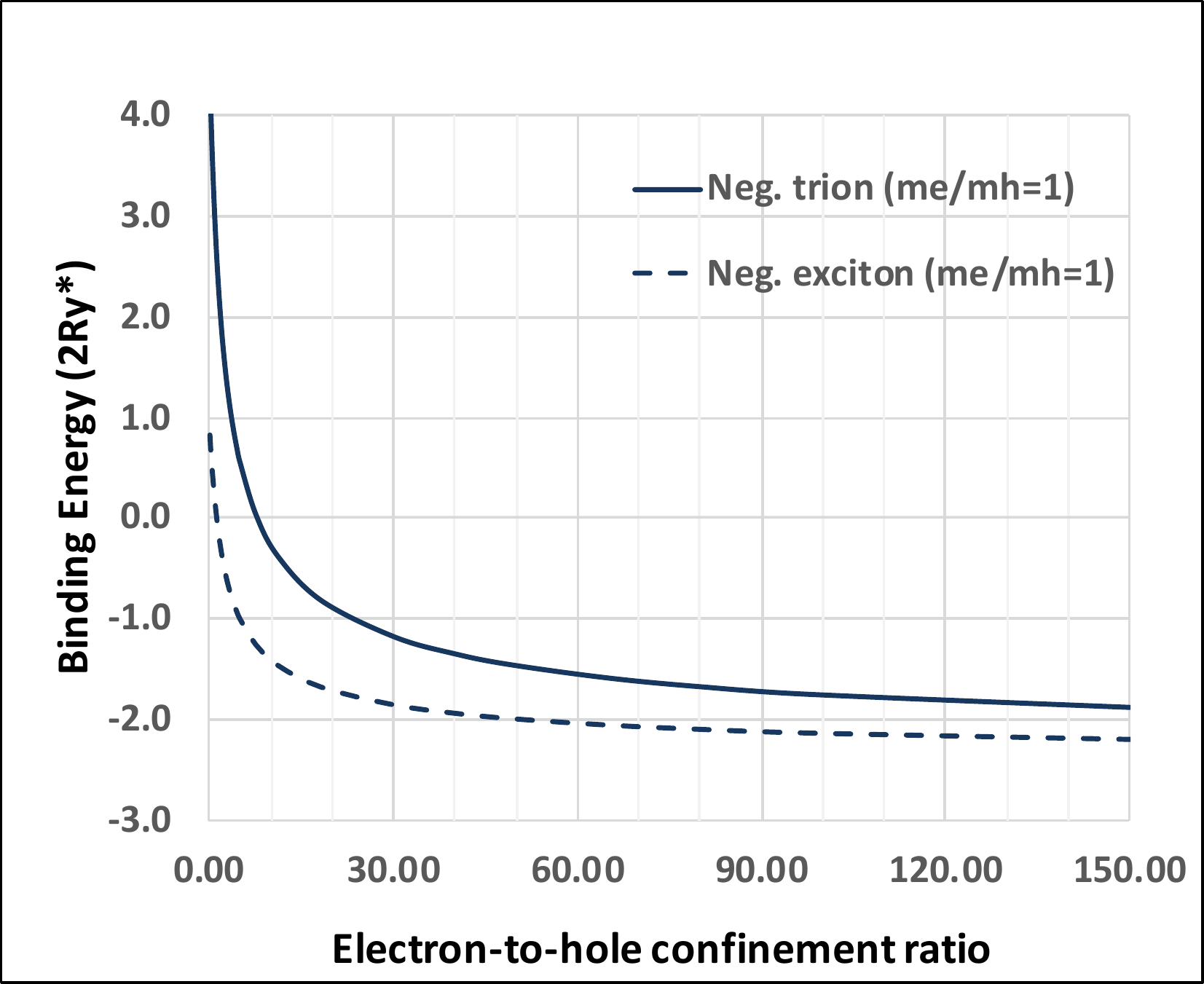}
\caption{\label{Fig4}(color online) The comparison of the negative trion and negatively charged exciton binding energies with the $m_e/m_h=1$ as function of electron-to-hole confinement ratio.}
\end{minipage} 
\end{center}
\end{figure}

In Figure \ref{Fig1}, Figure \ref{Fig2}, Figure \ref{Fig3} and Figure \ref{Fig4} the binding energies of the negatively charged exciton and the negative trion in the ground state without magnetic fields have been shown as  function  of  the confinement ratio between the electron and the hole. To see the impact of masses of carriers, we compare the results calculated for both parameter sets with $m_e/m_h =0.27$, and  $m_e/m_h =1$, respectively. 
One can see in the Figure \ref{Fig3} and Figure \ref{Fig4} that for the entire range of the electron-to-hole confinement ratio $\omega_e/\omega_h$, the negative trion binding energy is always larger than that of the negatively charged exciton. We notice that for small values of the electron-to-hole confinement ratio both negative trions and negatively charged excitons are bound systems but they become unbound when the confinement ratio between the electron and the hole increases. 

\begin{figure}[htbp]
\begin{center}
\begin{minipage}{6cm}
\includegraphics[width=6cm]{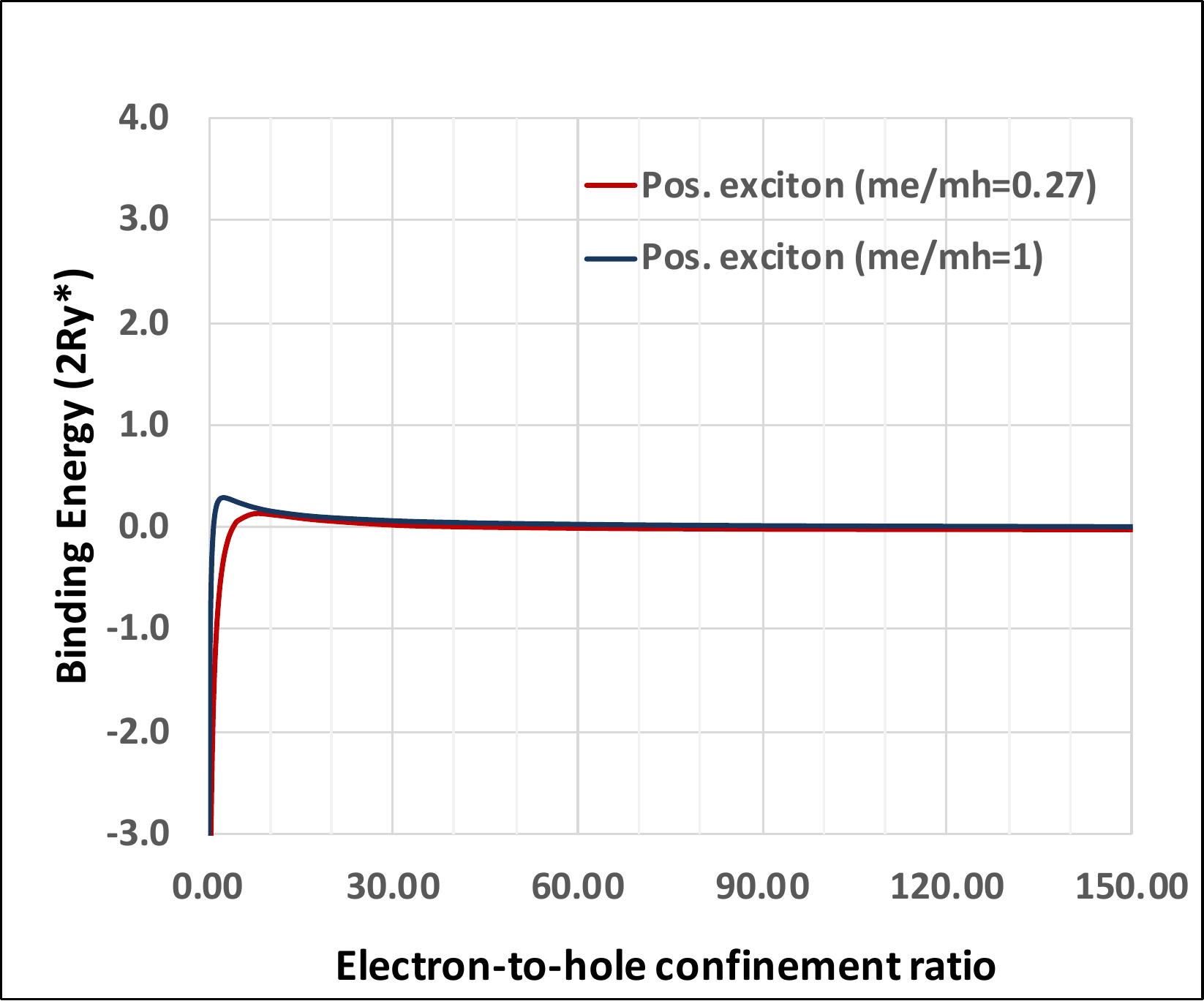}
\caption{\label{Fig5}(color online) The comparison of the positively charged exciton binding energies with  the $m_e/m_h=0.27$ and $m_e/m_h=1$ as function of electron-to-hole confenement ratio.}
\end{minipage}\hspace{2pc}%
\begin{minipage}{6cm}
\includegraphics[width=6cm]{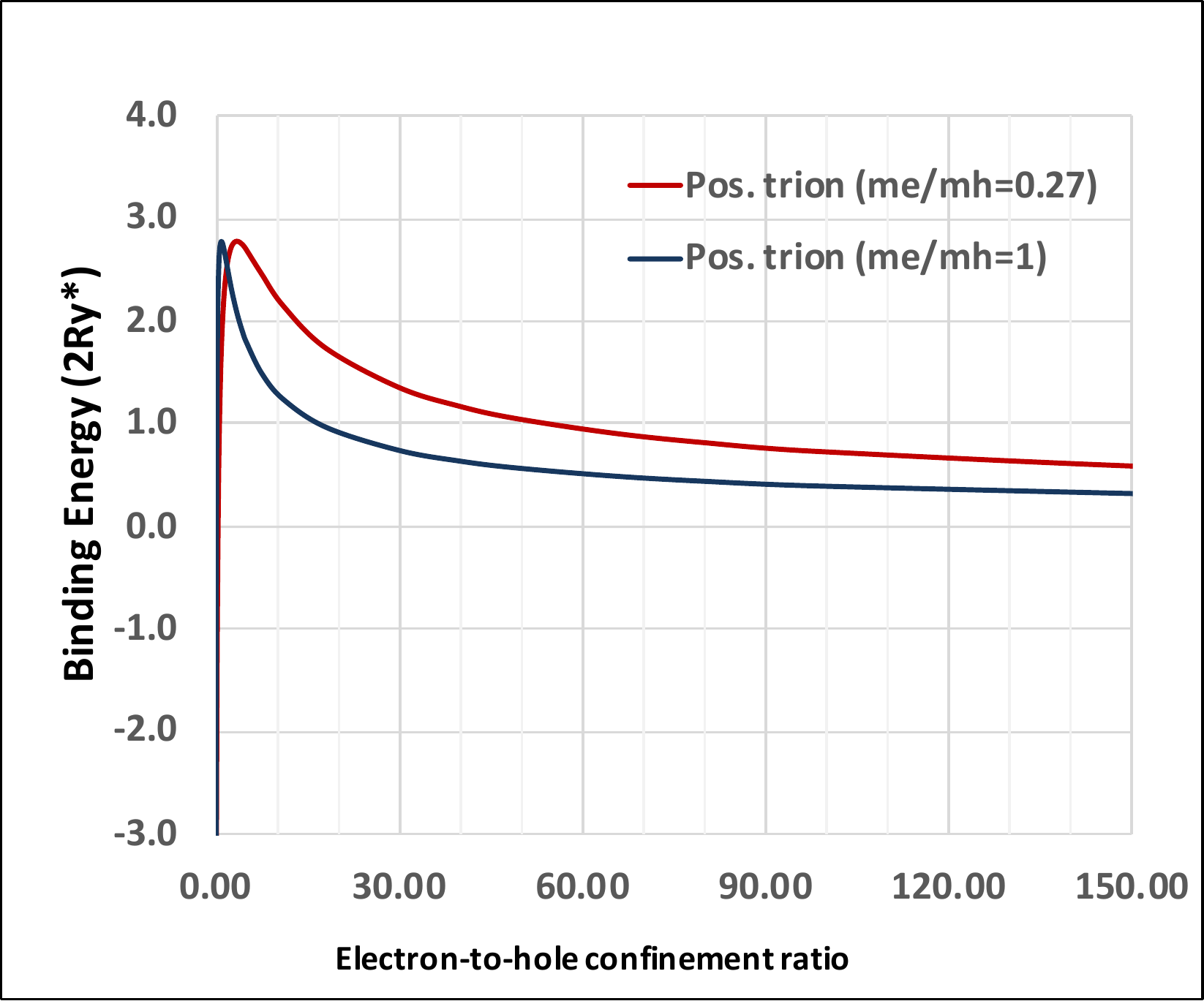}
\caption{\label{Fig6}(color online) The comparison of the positive trion binding energies with  the $m_e/m_h=0.27$ and $m_e/m_h=1$ as function of electron-to-hole confinement ratio.\\}
\end{minipage} 
\end{center}
\end{figure}

For the positively charged excitons and positive trions we also get the similar differences as between negatively charged excitons and negative trion. In Figure \ref{Fig5} and Figure \ref{Fig6} the binding energies of the positively charged excitons and positive trions in the ground states as function of electron-to-hole confinement ratio are compared in two parameter sets with $m_e/m_h=0.27$ and $m_e/m_h=1$. Again the positive trion binding energy is always larger than that of the positively charged excitons. 
\begin{figure}[h!]
\begin{center}
\begin{minipage}{6cm}
\includegraphics[width=6cm]{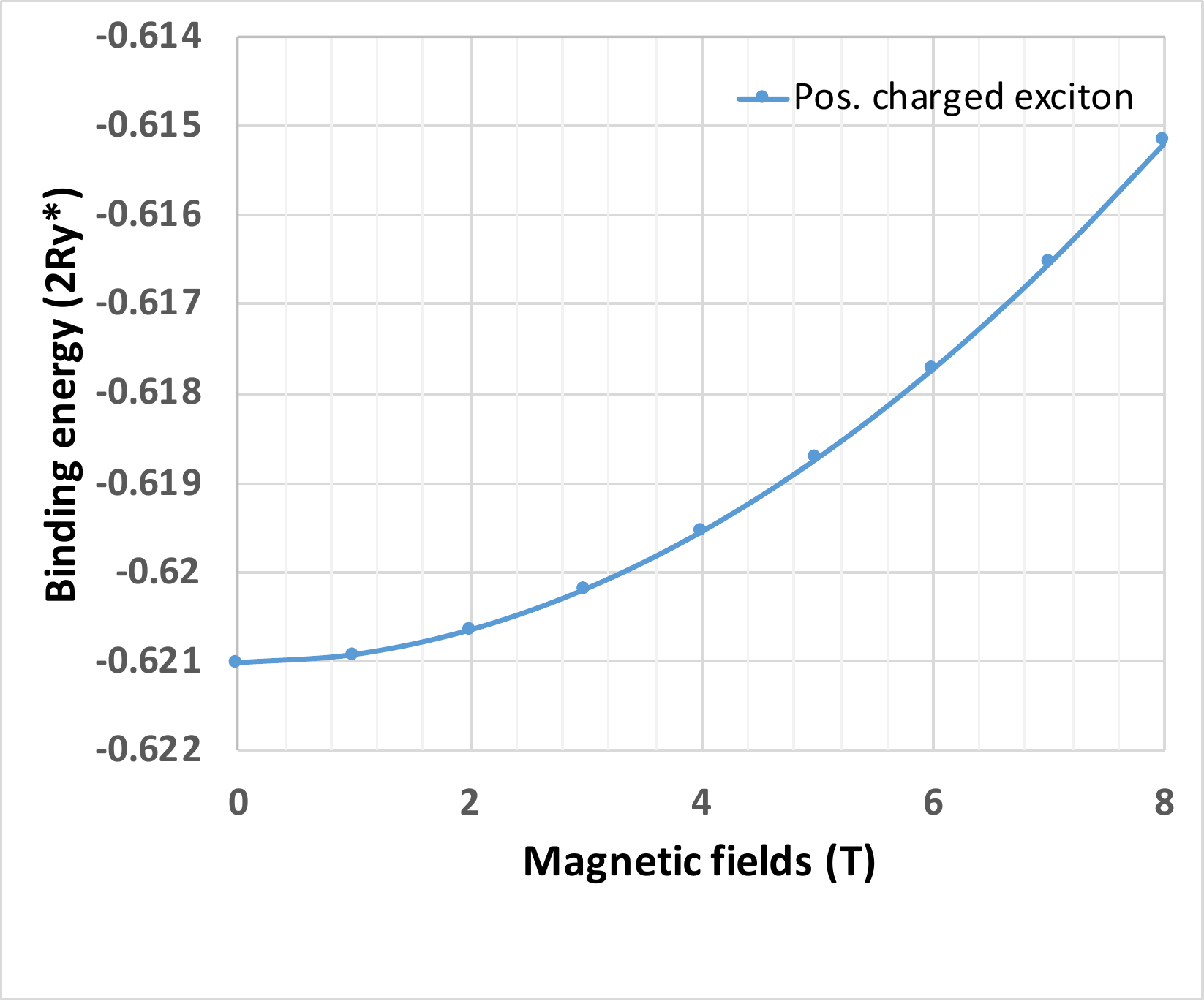}
\caption{\label{Fig7}(color online) The change of the negatively charged exciton binding energy as function of magnetic fields.}
\end{minipage}\hspace{2pc}%
\begin{minipage}{6cm}
\includegraphics[width=6cm]{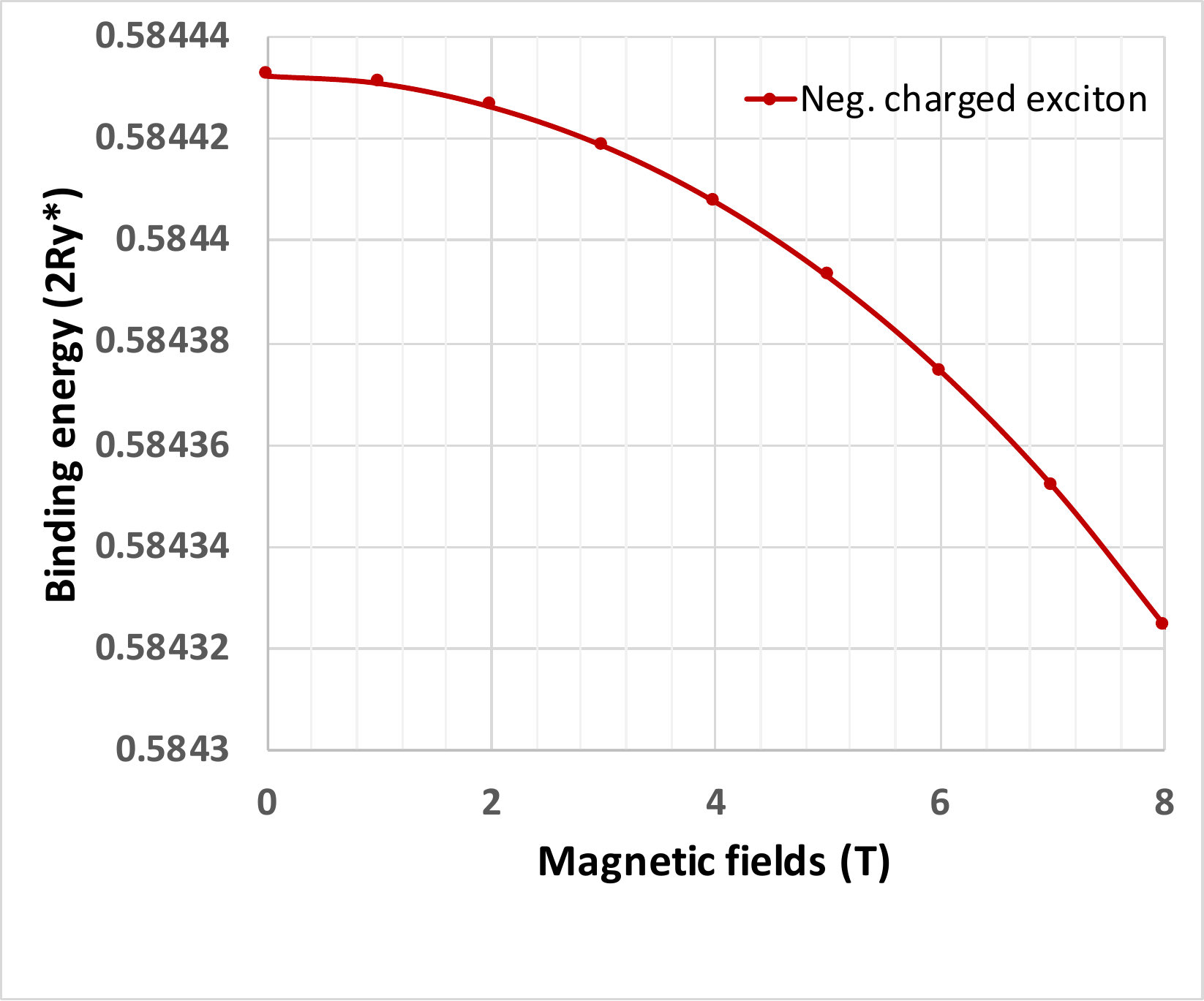}
\caption{\label{Fig8}(color online) The change of the positively charged exciton binding energy as function of magnetic fields.}
\end{minipage} 
\end{center}
\end{figure}
However, here we observe the transition from anti-binding to binding state of both positive trion and positively charged exciton when increasing the electron-to-hole confinement ratio. This situation is opposite for the above case of negative trions and negatively charged excitons, because the contributions of the hole-hole interaction is larger than that of electron-electron interaction to suppress the electron-hole interaction. 
It is interesting to note that our new results can help to understand and clarify sensitive changes in the binding energy of trions in natural ensembles of InAs/GaAs quantum dots with randomly fluctuating parameters \cite{Zielinski}. 

Figure \ref{Fig7} and Figure \ref{Fig8} show how the magnetic field affect the charged exciton state. It is seen that magnetic fields increase the binding energy of positively charged excitons but decrease it for the case of negatively charged exciton, although the effect is rather small. 

In \cite{Quang2,Quang3} we have analyzed the agreement between our results calculated by Hartree-Fock method and the experimental  data for the biexciton binding energy  \cite{Tsai}. In this work we also see that in InAs/GaAs self-assembled quantum dot we have the good agreement between our results and the experimental  data for the negatively and positively charged exciton binding energies too. 
Indeed, the value of electron-to-hole confinement ratio in InAs/GaAs self-assembled quantum dot is $\omega_e/\omega_h=1.96$, that corresponds to the value of the negatively and positively charged exciton binding energies $0.58\ (2Ry^*)\approx 6.73$ meV and $-0.62\ (2Ry^*)\approx -7.2$ meV, respectively. Our results agree very well with the experimental value of $6.2 \pm 0.4$ meV for the negatively exciton and rather good with the experimental value interval of [-1 meV:-6 meV]  for positively charged exciton, respectively \cite{Tsai}.
\section{Conclusion}
In conclusion, the charged exciton and trion states in 2D parabolic quantum dots have been studied in the Hartree-Fock-Roothaan formulation. The trion and charged exciton binding energies have been calculated for the entire range of the confinement ratio between the electron and the hole. It is shown that the binding energies of charged excitons and trions can be either negative or positive, depending on the correlation ratio of the electron-hole confinement potential, but binding energies of negative or positive trion systems are generally larger than that of the corresponding negatively or positively charged excitons.
Our theoretical calculation results are in good agreement with the experimental data on the binding energies of negative and positive trions in self-organizing small quantum dots InAs/GaAs.
It is shown that the binding energy of positively charged excitons is increased by magnetic fields while it is decreased in the case of negatively charged exciton, although the effect is rather small.  
Our results help further clarify the properties of the elementary excitations and the differences between charged excitons and trions in semiconductor quantum dots.

\section*{Acknowledgments}
This work is supported by the Vietnam Academy of Science and Technology (VAST) under project No. NVCC05.12/21-21 and partly by the project No.ICP.2021.01 of the International Centre of Physics (ICP), Institute of Physics.

\newcommand{\newblock}{}

\end{document}